\documentclass[twocolumn,english,aps,prl,reprint,superscriptaddress]{revtex4.1}
\usepackage[T1]{fontenc}
\usepackage[latin9]{inputenc}
\pdfoutput=1
\usepackage{amsmath}
\usepackage{amssymb}
\usepackage{graphicx}
\usepackage{subscript}

\makeatletter
\@ifundefined{textcolor}{}
{%
 \definecolor{BLACK}{gray}{0}
 \definecolor{WHITE}{gray}{1}
 \definecolor{RED}{rgb}{1,0,0}
 \definecolor{GREEN}{rgb}{0,1,0}
 \definecolor{BLUE}{rgb}{0,0,1}
 \definecolor{CYAN}{cmyk}{1,0,0,0}
 \definecolor{MAGENTA}{cmyk}{0,1,0,0}
 \definecolor{YELLOW}{cmyk}{0,0,1,0}
}

\usepackage{pslatex}

\usepackage{babel}

\usepackage{babel}

\usepackage{babel}

\makeatother

\usepackage{babel}
\begin{document}

\title{Sisyphus Laser Cooling of a Polyatomic Molecule}

\author{Ivan Kozyryev}

\email{ivan@cua.harvard.edu}

\affiliation{Harvard-MIT Center for Ultracold Atoms, Cambridge, MA 02138, USA}

\affiliation{Department of Physics, Harvard University, Cambridge, MA 02138, USA}

\author{Louis Baum}

\affiliation{Harvard-MIT Center for Ultracold Atoms, Cambridge, MA 02138, USA}

\affiliation{Department of Physics, Harvard University, Cambridge, MA 02138, USA}

\author{Kyle Matsuda}

\affiliation{Harvard-MIT Center for Ultracold Atoms, Cambridge, MA 02138, USA}

\affiliation{Department of Physics, Harvard University, Cambridge, MA 02138, USA}

\author{Benjamin L. Augenbraun }

\affiliation{Harvard-MIT Center for Ultracold Atoms, Cambridge, MA 02138, USA}

\affiliation{Department of Physics, Harvard University, Cambridge, MA 02138, USA}

\author{Loic Anderegg }

\affiliation{Harvard-MIT Center for Ultracold Atoms, Cambridge, MA 02138, USA}

\affiliation{Department of Physics, Harvard University, Cambridge, MA 02138, USA}

\author{Alexander P. Sedlack }

\affiliation{Harvard-MIT Center for Ultracold Atoms, Cambridge, MA 02138, USA}

\affiliation{Department of Physics, Harvard University, Cambridge, MA 02138, USA}

\author{John M. Doyle}

\affiliation{Harvard-MIT Center for Ultracold Atoms, Cambridge, MA 02138, USA}

\affiliation{Department of Physics, Harvard University, Cambridge, MA 02138, USA}
\begin{abstract}
We perform magnetically-assisted Sisyphus laser cooling of the triatomic
free radical strontium monohydroxide (SrOH). This is achieved with
principal optical cycling in the rotationally closed $P\left(N''=1\right)$
branch of either the $\tilde{X}^{2}\Sigma^{+}\left(000\right)\leftrightarrow\tilde{A}^{2}\Pi_{1/2}\left(000\right)$
or the $\tilde{X}^{2}\Sigma^{+}\left(000\right)\leftrightarrow\tilde{B}^{2}\Sigma^{+}\left(000\right)$
vibronic transitions. Molecules lost into the excited vibrational
states during the cooling process are repumped back through the $\tilde{B}\left(000\right)$
state for both the $\left(100\right)$ level of the Sr-O stretching
mode and the $\left(02^{0}0\right)$ level of the bending mode. The
transverse temperature of a SrOH molecular beam is reduced in one
dimension by two orders of magnitude to $\sim700\ {\rm \mu K}$. This
approach opens a path towards creating a variety of ultracold polyatomic
molecules, including much larger ones, by means of direct laser cooling. 
\end{abstract}
\maketitle
Compared to atoms, the additional rotational and vibrational degrees
of freedom in molecules give rise to a wide variety of potential and
realized scientific applications, including quantum computation \cite{demille2002quantum,rabl2006hybrid,andre2006coherent},
precision measurements \cite{ACME2014,hilborn1996spectroscopic,demille2015diatomic,demille2008using},
and quantum simulation \cite{micheli2006toolbox,gorshkov2013topological}.
While ultracold diatomic molecules will be extremely valuable in opening
novel research frontiers, molecules with three or more atoms have
unique capabilities enabled by their significantly more complicated
structure \cite{pastor2015testing,kozlov2013linear,shelkovnikov2008stability,wall2015realizing,wall2013simulating,quack2002important,quack2008high}.
For all molecules to achieve their full scientific potential, they
must be cooled. Yet, the desired quantum complexity that molecules
provide also leads to challenges for control, detection, and cooling
\cite{carr2009cold}. Assembling ultracold molecules from two laser-cooled atoms has represented one solution and has created ultracold
bi-alkali molecules \cite{ni2008high,park2015ultracold,takekoshi2014ultracold,molony2014creation,guo2016creation},
including filling of optical lattices with KRb \cite{moses2015KRb}.
There are several direct cooling techniques that together routinely
cool a much wider variety of molecules into the Kelvin regime \cite{lemeshko2013manipulation,carr2009cold}.
Intense research is ongoing to bring these cold molecules into the
ultracold regime ($<1\ {\rm mK}$). Even though there has been experimental
progress on control of polyatomics \cite{zeppenfeld2012sisyphus,chervenkov2014continuous,bethlem2000electrostatic,patterson2015slow,momose2013manipulation,fulton2004optical},
optoelectrical cooling of formaldehyde is the only technique that
has resulted in a trapped sub-millikelvin sample \cite{prehn2016optoelectrical}.

Cooling of the external motion of neutral atoms from above room temperature
into the sub-millikelvin range (leading to, e.g., Bose-Einstein condensation)
commonly relies on the use of velocity-dependent optical forces \cite{adams1997laser}.
Laser cooling requires reasonably closed and strong optical electronic
transitions, so its use for molecules has been severely limited. Recently,
following initial theoretical proposals \cite{di2004laser,stuhl2008magneto}
and proof-of-principle experimental results \cite{Shuman2009}, laser
cooling has been achieved for SrF \cite{Shuman2010}, YO \cite{Hummon2013},
and CaF \cite{zhelyazkova2014laser,hemmerling2016CaF}, including
a magneto-optical trap for SrF \cite{barry2014magneto,mccarron2015improved,norrgard2015sub}.
Motivated by this progress on diatomic molecules, and building upon
previous theoretical work \cite{isaev2015polyatomic}, we recently
demonstrated photon cycling -- a crucial requirement for achieving
light induced forces -- with the triatomic molecule SrOH \cite{kozyryev2016radiation}.

In this Letter, we report the Sisyphus laser cooling of a polyatomic
molecule. The dissipative force for compressing phase-space volume
is achieved by a combination of spatially varying light shifts and
optical pumping into dark sub-levels, which are then remixed by a
static magnetic field, as explored previously in atomic systems \cite{emile1993magnetically,sheehy1990magnetic}.
Since the magnitude of the induced friction force is directly related
to the modulation depth of the dressed energy levels, the cooling
process can be much more efficient than with Doppler radiation pressure
forces \cite{aspect1986cooling,padua1993transient}. This enhancement
is especially important for complex polyatomic molecules, where scattering
the thousands of photons necessary for Doppler cooling becomes more
challenging due to additional vibrational modes. Here, we demonstrate
transverse cooling (and heating) of a SrOH beam using two different
electronic transitions, study loss channels to vibrational states
(including the bending mode), and highlight proposed extensions to
more complex species.

\begin{figure}[h]
\begin{centering}
\includegraphics[width=12cm]{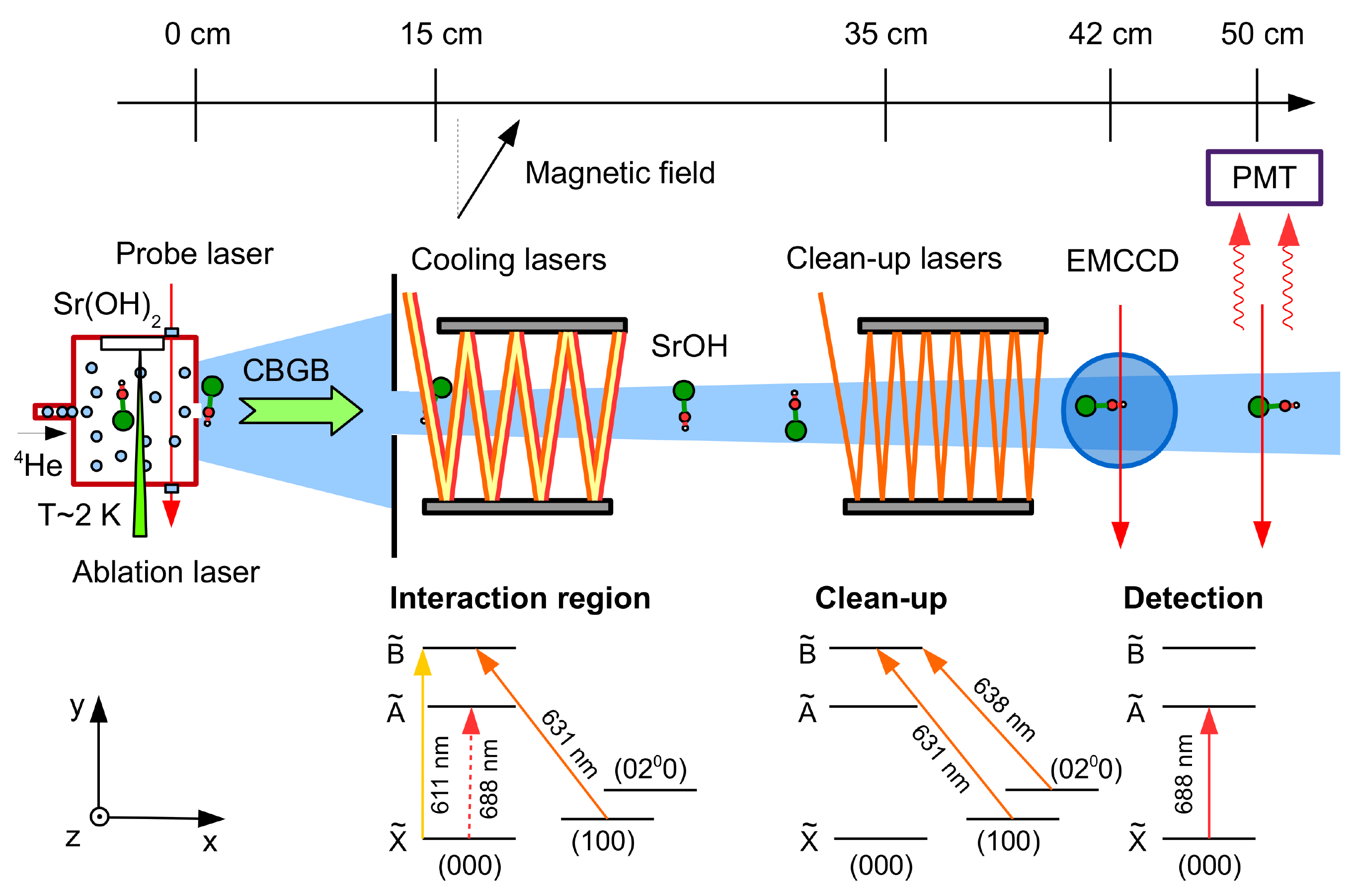} 
\par\end{centering}

\protect\protect\protect\protect\caption{\label{fig:Schematic-of-the-apparatus}Schematic of the experimental
apparatus (not to scale). A cryogenic beam of SrOH is produced using
laser ablation of a pressed Sr(OH)\protect\protect\protect\textsubscript{2}
target followed by buffer-gas cooling with $\sim2\ {\rm K}$ helium
gas. To apply the cooling forces on the collimated molecular beam,
we use transverse lasers retroreflected between two mirrors in order
to generate a standing wave. All of the lasers are resonant with the
$P\left(N''=1\right)$ rotationally-closed line of the corresponding
electronic transition. Depending on the experimental configuration,
either the $\tilde{X}^{2}\Sigma^{+}\left(000\right)\rightarrow\tilde{A}^{2}\Pi_{1/2}\left(000\right)$
or the $\tilde{X}^{2}\Sigma^{+}\left(000\right)\rightarrow\tilde{B}^{2}\Sigma^{+}\left(000\right)$
cooling transition is used with an additional $\tilde{X}^{2}\Sigma^{+}\left(100\right)\rightarrow\tilde{B}^{2}\Sigma^{+}\left(000\right)$
laser for repumping molecules decaying to the vibrationally excited
Sr-O stretching mode. In order to remix dark magnetic sub-levels,
a magnetic field is applied in the interaction region. Before the
detection is performed, molecules remaining in either $\left(100\right)$
or $\left(02^{0}0\right)$ excited vibrational levels of the electronic
ground state are optically pumped back into the ground vibrational
level using $\tilde{X}\rightarrow\tilde{B}$ off-diagonal excitations.
The spatial profile of the molecular beam is imaged on the electron
multiplying charge-coupled device (EMCCD) camera and the time-of-flight
(ToF) data is collected on the photomultiplier tube (PMT). The vibrational
quantum numbers $\left(v_{1}v_{2}^{l}v_{3}\right)$ correspond to
the Sr$\leftrightarrow$OH stretching ($v_{1}$), Sr-O-H bending ($v_{2}$),
and SrO$\leftrightarrow$H stretching ($v_{3}$) vibrational modes.
The superscript $l$ next to the bending mode vibrational quantum
number indicates the projection of the vibrational angular momentum
on the internuclear axis. }
\end{figure}

Our work with SrOH uses the cryogenic buffer-gas beam (CBGB) \cite{hutzler2012buffer},
which is also used in all other experiments on laser cooling of molecules.
The study of SrOH buffer-gas cooling dynamics, as well as precise
measurements of its momentum transfer and inelastic cross sections
with helium, were previously performed \cite{kozyryev2015collisional}.
In brief, SrOH can be produced efficiently with ablation and forms
an intense CBGB. Fig. \ref{fig:Schematic-of-the-apparatus} shows
a simplified schematic diagram of the current experimental apparatus.
Detailed descriptions of this apparatus have also been provided elsewhere
\cite{hemmerling2013buffer}. Laser ablation of Sr(OH)\textsubscript{2}
produces SrOH molecules that are then entrained in helium buffer gas
($T_{{\rm He}}\sim2\ {\rm K}$) that flows out of the cell into a
beam. He flow is 6 standard cubic centimeters per minute (sccm), and
the beam is extracted through a $5\ {\rm mm}$ diameter aperture.
This CBGB contains $\sim10^{9}$ molecules in the first excited rotational
level ($N=1$) in a pulse $\sim5\ {\rm ms}$ long. The forward velocity
of the SrOH beam is $v_{x}\sim130\ {\rm m/s}$ and its transverse
velocity spread is $\triangle v_{y}\sim\pm15\ {\rm m/s}$. A
$2\times2$ mm square aperture situated 15 cm away from the cell collimates
the beam, resulting in an effective transverse temperature $T_{\perp}\sim50\ {\rm mK}$.

To laser cool, we use a photon cycling scheme that we also employed
in an earlier work, as described in detail in Ref. \cite{kozyryev2016radiation}.
The main photon cycling path is $\tilde{X}^{2}\Sigma^{+}\left(000\right)\rightarrow\tilde{B}^{2}\Sigma^{+}\left(000\right)$
(611 nm) and the first vibrational repump is $\tilde{X}^{2}\Sigma^{+}\left(100\right)\rightarrow\tilde{B}^{2}\Sigma^{+}\left(000\right)$
(631 nm), as shown in Fig. \ref{fig:Schematic-of-the-apparatus} (Interaction
region). The combined main and repump laser light, with diameter of
$\sim3\ {\rm mm}$, propagates in the $y$ direction and makes 5 round-trip
passes between two mirrors before it is retroreflected back in order
to create a standing wave. The molecule-laser interaction time is
$t_{{\rm int}}\sim115\ {\rm \mu s}$. Each color (611 nm and 631 nm) includes two frequency components separated
by $\sim110\ {\rm MHz}$ to address the $P_{11}\left(J''=1.5\right)$
and $^{P}Q_{12}\left(J''=0.5\right)$ lines of the spin-rotation (SR)
splitting. We also study cooling using the $\tilde{X}^{2}\Sigma^{+}\left(000\right)\rightarrow\tilde{A}^{2}\Pi_{1/2}\left(000\right)$
excitation at 688 nm as the main transition. Each SR component of
the 688 nm light is generated using separate injection-locked laser
diodes seeded by external-cavity diode lasers in the Littrow configuration
\cite{cunyun2004tunable} resulting in $\sim15\ {\rm mW}$ per SR
component in the interaction region. The 611 nm light, as well as
all of the repumping light, is generated by cw dye lasers passing
through acousto-optic modulators resulting in $\sim50\ {\rm mW}$
per SR component. In order to destabilize dark states created during
the cycling process \cite{berkeland2002destabilization}, we apply
a magnetic field of a few gauss. Due to the vibrational angular momentum
selection rule \cite{herzberg1966molecular}, the dominant loss channel
for the bending mode is to the $v_{2}=2$ state with $l=0$ \cite{OberlanderPhDthesis}
denoted $\left(02^{0}0\right)$. Further details regarding the photon
cycling scheme used for SrOH have been previously described \cite{kozyryev2016radiation}.

The spatial profile of the molecular beam is recorded by imaging laser-induced
fluorescence (LIF) in the Detection region. The molecules are excited
using a transverse retroreflected laser beam and LIF photons are imaged
onto an EMCCD camera. The detection laser addresses both SR components
of the $P\left(N''=1\right)$ line for the $\tilde{X}^{2}\Sigma^{+}\left(000\right)\rightarrow\tilde{A}^{2}\Pi_{1/2}\left(000\right)$
transition, as shown in Fig. \ref{fig:Schematic-of-the-apparatus}
(Detection). In a similar laser configuration, time of flight (ToF)
data is recorded by collecting the LIF on a PMT (further downstream).
In order to boost the LIF signal there is a Clean-up region where
all of the molecular population is pumped into the ground state ($\tilde{X}\left(000\right)$)
from the excited vibrational levels ($\tilde{X}\left(100\right)$
and $\tilde{X}\left(02^{0}0\right)$). This is done with off-diagonal
excitation to $\tilde{B}\left(000\right)$, as shown in Fig. \ref{fig:Schematic-of-the-apparatus}
(Clean-up).

\onecolumngrid

\twocolumngrid

\begin{figure}[h]
\begin{centering}
\includegraphics[width=16cm]{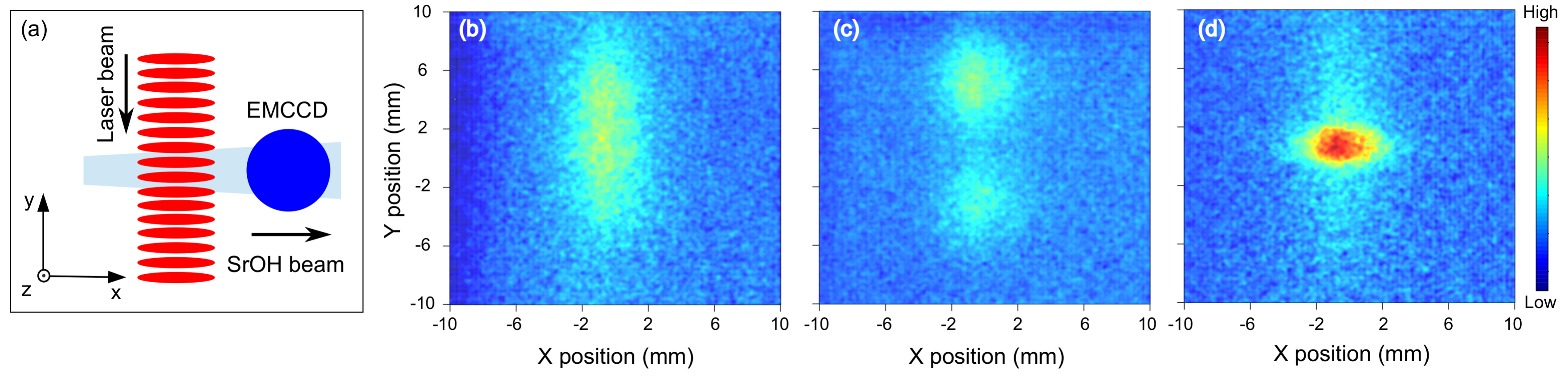} 
\par\end{centering}

\protect\protect\protect\protect\caption{\label{fig:610nm-2D-images}Spatial images of the molecular beam taken
at different detunings of the $\tilde{X}\left(000\right)-\tilde{B}\left(000\right)$
cooling laser: (b) on resonance, (c) red-detuned (-10 MHz), and (d)
blue-detuned (+10 MHz). The same color axis is used for all three
plots. SrOH beam is moving in the $x$ direction while the cooling
laser is applied in the $y$ direction as shown in (a). Narrowing
of the spatial size of the molecular cloud with accompanying density
increase in (d) compared to (b) in the $y$ dimension indicates phase-space
compression. }
\end{figure}

Fig. \ref{fig:610nm-2D-images} shows 2D camera images of the molecular
beam for various detunings of the $\tilde{X}-\tilde{B}$ cooling laser. Phase-space compression is clearly visible in the comparison
between images (b), $\delta=0$, and (d), $\delta>0$, cooling.

\begin{figure}[h]
\begin{centering}
\includegraphics[width=12cm]{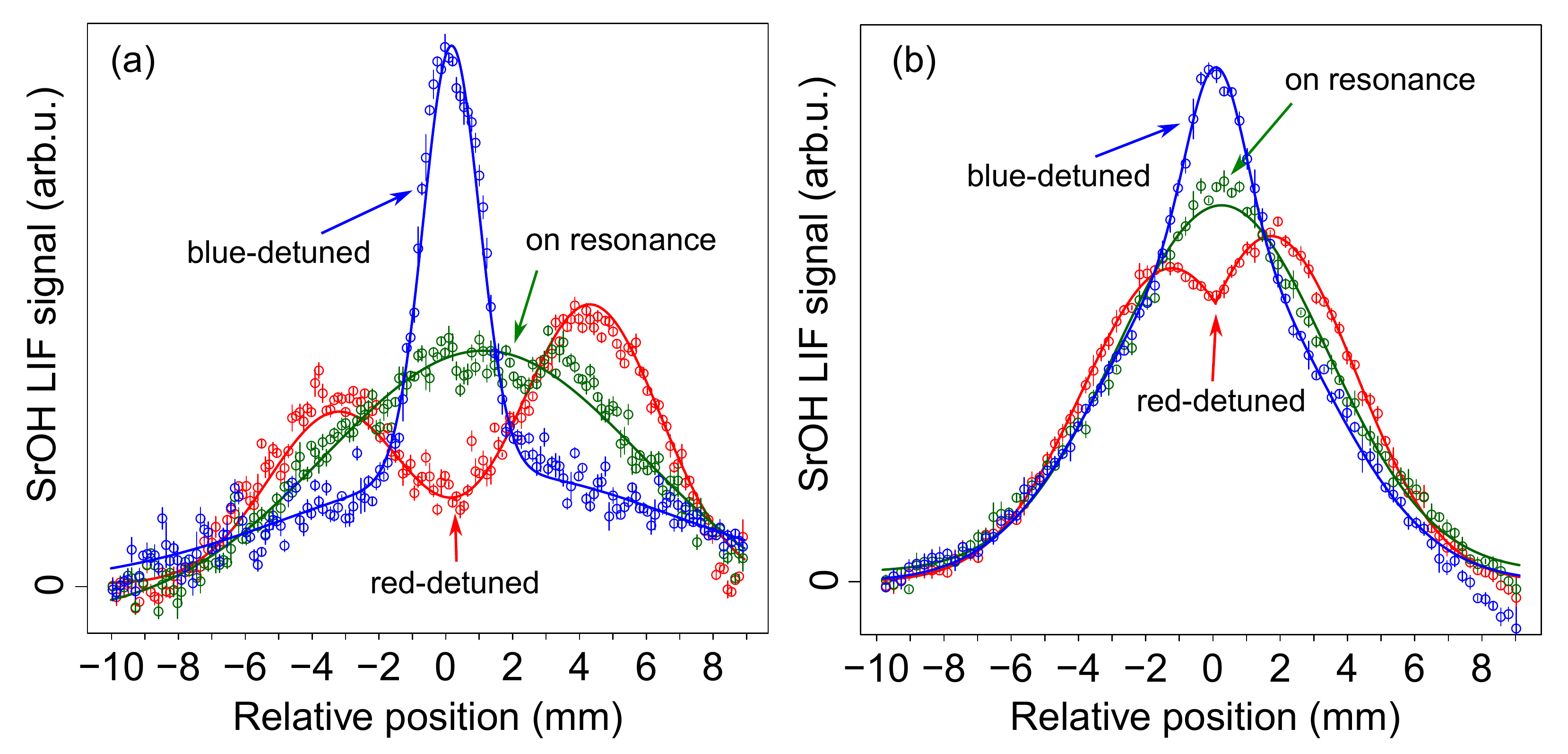} 
\par\end{centering}

\protect\protect\protect\protect\caption{\label{fig:688nm-cooling}Integrated molecular beam profiles for different
detunings of the cooling laser: (a) $\tilde{X}\left(000\right)-\tilde{B}\left(000\right)$
and (b) $\tilde{X}\left(000\right)-\tilde{A}\left(000\right)$. The
detunings from resonance are given by $\delta=\pm10\ {\rm MHz}$.
With a positive detuning, the width of the molecular beam is reduced,
which indicates cooling of the molecular beam: (a) $9.4\pm0.3\ {\rm mm}$
to $1.67\pm0.03\ {\rm mm}$ and (b) $6.3\pm0.1\ {\rm mm}$ to $2.1\pm0.1\ {\rm mm}$.
A ``hole'' around zero for $\delta<0$ represents a heating signature
of the magnetically-assisted Sisyphus effect with the widening of
the total spatial distribution. An asymmetry in the height of two
peaks comes from imperfect alignment between laser and molecular beams
and was previously seen in similar experiments with atoms \cite{aspect1986cooling}.
The excess signal above the fit near zero position for the on resonance
trace in (b) is potentially indicative of a slightly positive detuning
of the cooling lasers. }
\end{figure}

To characterize the cooling efficacy for both $\tilde{X}-\tilde{A}$
and $\tilde{X}-\tilde{B}$ cycling transitions, we plot integrated
1D ($x$ axis) beam profiles for both cooling configurations in Fig.
\ref{fig:688nm-cooling}. The most effective laser cooling was demonstrated
using $\tilde{X}\left(000\right)-\tilde{B}\left(000\right)$ transition
at 611 nm with laser intensity $I=1.4\ {\rm W/cm^{2}}$, resulting
in a saturation parameter $s\sim20$ (Fig. \ref{fig:688nm-cooling}(a)).
From the fits to the data and a comparison to Monte Carlo (MC) simulations
of the molecular beam kinetics we determine the final beam temperature
$T_{\perp}\sim700\ {\rm \mu K}$, which corresponds to a factor of
70 reduction as compared to the $\delta=0$ detuning. Because of the
high damping rate, we achieve lower transverse temperature than previously
demonstrated with a 1D MOT of diatomic molecules \cite{Hummon2013},
with half the interaction length.

Cooling using the $\tilde{X}-\tilde{A}$ transition was less effective.
Fig. \ref{fig:688nm-cooling}(b) shows typical molecular beam profiles
after interacting with a cooling laser exciting the $\tilde{X}-\tilde{A}$
transition at 688 nm with intensity of $I=424\ {\rm mW/cm^{2}}$ and
a saturation parameter $s\sim8$. For a positive detuning we observe
cooling of the SrOH beam represented by the increased molecular density
near the center due to the narrowing of the spatial distribution.
By comparing the fitted width of the resulting profile with a MC simulation
we conclude that the beam is cooled to a final temperature of $\sim2\ {\rm mK}$,
an order of magnitude above the Doppler limit of $\sim200\ {\rm \mu K}$.

In order to extract the number of scattered photons during the cooling
process, we determine the fraction of the remaining molecules after
the cooling process with ToF PMT data taken without the $\left(02^{0}0\right)$
clean-up beam. Using the previously measured decay rate to dark vibrational
levels (above $\tilde{X}\left(100\right)$) of $\left(3\pm1\right)\times10^{-3}$
\cite{kozyryev2016radiation}, we calculate that on average each molecule
emits $220_{-60}^{+110}$ photons with a scattering rate of $\Gamma_{{\rm scat}}=2\pm1\ {\rm MHz}$.
In such a configuration, Doppler cooling from radiation-pressure molasses
does not play a significant role \cite{aspect1986cooling}. By adding
the $\left(02^{0}0\right)$ clean-up beam, we determine that $\sim10\%$
of molecules decay to the $\left(02^{0}0\right)$ state of the bending
mode during the cooling process.

\onecolumngrid

\twocolumngrid

For negative detunings the molecules are expelled from the region
around $v_{y}=0$, leading to a double-peak structure that is a signature
of the magnetically-assisted Sisyphus effect \cite{emile1993magnetically}.
Compared to the results of cycling on the $\tilde{X}-\tilde{A}$ transition
(Fig. \ref{fig:688nm-cooling}(b)), the use of the $\tilde{X}-\tilde{B}$
transition (Fig. \ref{fig:688nm-cooling}(a)) increases the separation
between the peaks from $2.95\pm0.04\ {\rm mm}$ to $7.54\pm0.04\ {\rm mm}$
for $\delta<0$. Our findings are in good agreement with previous
studies of sub-Doppler laser cooling in complex multilevel atomic
\cite{kloter2008laser} and molecular systems \cite{Shuman2010,devlin2016three}.

\enlargethispage{2\baselineskip}

In summary, we demonstrate Sisyphus laser cooling of the polyatomic
molecule SrOH. We reduce the transverse temperature of a cryogenic
buffer-gas beam from $50\ {\rm mK}$ to $700\ {\rm \mu K}$ with $\sim200$
scattered photons per molecule. Laser cooling of atoms is a mature
scientific field \cite{chu1998nobel,aspect1995laser,phillips1998nobel}
with well developed experimental \cite{aspect1992manipulation,metcalf2003laser}
and theoretical \cite{cohen1992laser,dalibard1989laser} techniques.
Our results with SrOH open up a wide range of future directions for
laser manipulation of polyatomic molecules. Cooling SrOH motion with
magnetically-assisted laser cooling close to the recoil temperature
of $\sim1\ {\rm \mu K}$ should be possible by increasing the interaction
time and optimizing laser power \cite{hoogerland1992magnetically,sheehy1990collimation}.
Extending the scheme to 2D and using more elaborate optical configurations
would lead to significantly increased brightening of the molecular
beam \cite{hoogerland1996bright,sheehy1990collimation}. Slowing and
cooling of an atomic beam in the longitudinal dimension \cite{soding1997short,prentiss1989slowing},
e.g. for loading into a MOT, could now be extended to polyatomic molecules.

While some of these research avenues might require repumping of other
vibrational states beyond the $\left(100\right)$ and $\left(02^{0}0\right)$
states as the number of scattered photons increases, this challenge
can be solved with additional repumping lasers on the $\tilde{X}-\tilde{B}$
transition. Since the strengths of higher-order Franck-Condon factors
decrease rapidly \cite{OberlanderPhDthesis,nicholls1981franck}, scattering
of $\sim10,000$ photons should be possible with only two additional
lasers to address $\left(200\right)$ and $(01^{1}0)$ states. All
of the required frequencies can be generated with solid-state laser
diodes that have easily attainable requisite powers \cite{ball2013high}.
Moreover, by using $\tilde{X}-\tilde{A}$ electronic excitation for
laser cooling and $\tilde{X}-\tilde{B}$ excitation for repumping,
the scattering rate becomes independent of the number of repumping
lasers, ensuring rapid optical cycling at a maximum possible rate.

While SrOH has a linear geometry in the vibronic ground state, it
still serves as a useful test candidate for the feasibility of laser
cooling more complex, nonlinear molecules like strontium monoalkoxide
free radicals, where hydrogen is replaced by a more complex group
R (e.g. R\,=\,CH\textsubscript{3}). Since SrOR molecules share
a number of properties with SrOH, including a very ionic Sr-O bond,
diagonal Franck-Condon factors, and technically accessible laser transitions
\cite{brazier1986laser,bernath1997spectroscopy}, results presented
in this paper could naturally be extended to such complex species
\cite{kozyryev2016MOR}.

\pagebreak{}


\onecolumngrid

\twocolumngrid

We would like to thank D. DeMille for insightful discussions. We would
also like to acknowledge contributions of B. Hemmerling to the earlier
stages of the experiment. This work was supported by the AFOSR.

 \bibliographystyle{apsrev}
\bibliography{Sisyphus_cooling_v1}

\end{document}